\documentclass[twocolumn,prl,aps]{revtex4}
\usepackage{graphicx}

\begin{document}

\title{Partially ferromagnetic electromagnet for trapping and cooling neutral atoms to quantum degeneracy}

\author{M. Fauquembergue,  J-F. Riou, W. Guerin, S. Rangwala\footnote{Present address: Raman Research Institute, C.V. Raman Avenue, Sadashivanagar, BANGALORE 560080 (INDIA)}, F. Moron, A. Villing, Y. Le Coq, P. Bouyer and \\ A. Aspect}

\affiliation{Groupe d'Optique Atomique\\ Laboratoire Charles Fabry
de l'Institut d'Optique\\ UMRA 8501 du CNRS\\ B\^{a}t. 503
Campus universitaire d'Orsay\\ 91403 ORSAY Cedex
(FRANCE)}
\author{M. L\'{e}crivain}
\affiliation{SATIE\\ Laboratoire de l'Ecole Normale Sup\'{e}rieure de Cachan \\ UMR 8029 du CNRS \\ 61, avenue du Pr\'{e}sident Wilson \\ 94235 CACHAN Cedex (FRANCE) }

\begin{abstract}
We have developed a compact partially ferromagnetic electromagnet to produce a Ioffe-Pritchard trap for neutral atoms. Our structure permits strong magnetic confinement with low power consumption. Compared to the previous iron-core electromagnet \cite{old electromagnet}, it allows for easy compensation of remnant fields and very high stability, along with cost-effective realization and compactness. We describe and characterize our apparatus and demonstrate trapping and cooling of $^{87}$Rb atoms to quantum degeneracy. Pure Bose-Einstein condensates containing $\mathrm{10^6}$ atoms  are routinely realized on a half-minute cycle. In addition we test the stability of the magnetic trap by producing atom lasers.
\end{abstract}

\date{\today}
\maketitle

\section{Introduction}

The
realization of Bose-Einstein
condensates (BECs) of a dilute gas of
trapped atoms \cite{firstBEC} has opened 
the way to great improvements in the domain of atom optics and atom interferometry. 
The
macroscopic population of atoms in a single quantum state
is the matter-wave
analog to a laser in optics \cite{Mewes97, Anderson98, Hagley99, Bloch99} for which the generalization of
its use has revolutionized interferometry technology \cite{Cho85, Ste95, Stedman97}. It is therefore expected that the use
of Bose-Einstein condensed atoms will advance the field of atom
optics and non linear atom optics \cite{nonlinearatopt}. In particular, BECs will bring about an
unprecedented level of accuracy in atom interferometry  \cite{Bouyer97,Gupta02,andrews}.

In the several years since their  first observation, BECs have been studied
 extensively \cite{ReviewBEC}. In most cases the condensate
properties themselves are the focus of the investigations. Relatively
little work has been done so far using a condensate as a tool to explore questions
in other  fields \cite{Gupta02,hmlecoq,casimircornell}. Therefore, developing a system that
could be used for these purposes is worthwhile, but needs a special focus in the 
conception of the key elements. 
One of the main challenges when designing such a BEC apparatus is to produce in a stable and reproducible way Bose-Einstein condensates with a large number of atoms.
For that, one needs to optically collect many atoms and have a long lifetime for
the atoms in the trap where the atoms will be cooled down to a few hundred nanokelvins. 
To reach these ultra low temperatures and enter the quantum degeneracy regime, evaporative cooling of 
trapped atoms \cite{Hess} 
remains at this time the most efficient method. This technique relies on a sufficiently high binary 
collision rate in order to allow fast thermalization of the external degrees of freedom of the atoms \cite{Walraven}.
Thus, besides requiring an ultrahigh vacuum and a sophisticated
timing system, the major challenge lies in designing an
adequate and easy-to-use non-dissipative trap with high confinement, either using magnetic or optical fields, 
which forms the
core element of each BEC experiment.
Potential applications, such as inertial sensors in space \cite{hmlecoq}, mandate the 
highest gradients for the lowest power.

Several groups around the world have unveiled
condensate-producing technologies that may in the future
prove to fullfill the requirements.

One such system involves ``macroscopic" magnets, 
either permanent or electrically controlled \cite{lewandowski,QUIC,old electromagnet,hulet}.
A magnetic  configuration commonly used is the Ioffe-Pritchard trap configuration \cite{IoffePritchard} because 
it avoids trap losses due to nonadiabatic spin flips: a quadrupolar linear trap providing the radial confinement is 
combined with a dipolar trap providing the axial confinement. To increase the density of trapped atoms and 
ensure a high collision rate, tight confinement is necessary. To accomplish those requirements, Ioffe traps typically dissipate kilowatts of power necessitating considerable cooling and stabilization as well as causing electronic switching problems. 
Different structures of magnetic traps have been widely used and 
a few, such as the QUIC traps \cite{QUIC}, the ferromagnetic traps \cite{old electromagnet}, the permament magnets \cite{hulet}
or a combination \cite{lewandowski} have very low power consumption.
Another system called ``atom chip" involves a surface magneto-optical trap (MOT)
 and a magnetic trap based on a wafer with lithographically patterned wires \cite{Puce}. This technology is compact, generates condensates
with unprecedented rapidity, and holds promise for eventually being
simpler and more robust than traditional condensate machines. Nevertheless
it requires more expertise to fabricate \cite{rugositechip} and 
might not allow for high fluxes because of its small size. 
The latest approach is the all optical method \cite{OpticalBEC}. 
By eliminating the magnetic trap altogether, this method indeed
may eventually become the simplest route to BEC, but the need for high power lasers makes it
still difficult with the present technology.

Our system is a hybrid magnetic trap that falls in the first category. It uses
strong iron core magnets to produce radial confining  fields and low power
electromagnetic coils to produce axial confinement and a bias  field. The ferromagnets guide and concentrate the magnetic flux where it is needed, i.e. on the atoms, and thus can be used to create a compact and light aparatus that consumes very little power. They are used above saturation threshold and
thus, do not need to be actively temperature controlled like permanent magnets \cite{hulet} and only need moderate cooling compared to electromagnetic coils producing the same  field. The tight
confinement from the ferromagnets insures that we have the necessary
collision rate to evaporatively cool the atoms.  The longitudinal bias  field, the only parameter for
which stability is critical, is produced by servo-controlled electromagnetic
coils, which are water cooled.

In this paper, we report on the development of this new hybrid electromagnet.
In the first section we explain the main advantages of ferromagnetic trapping. In the second and third section we describe this electromagnet and 
present the way it is used to cool $^{87}$Rb atoms.
In the last section we present stability measurements made using atom lasers.

\section{1. Why ferromagnetic trapping?} 

\begin{figure}[htb]  
\centering
  \resizebox{0.9\columnwidth}{!}{\includegraphics{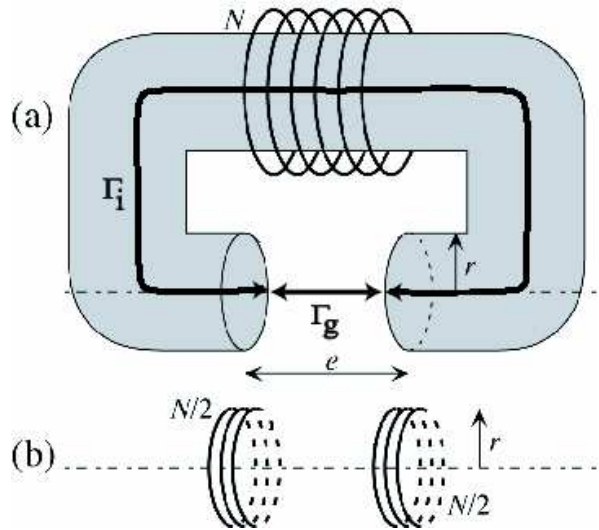}}  
\caption{(a) Ferromagnetic structure excited by a coil made of $N$ loops, with a gap $e$ between the north pole and the south pole. (b) Equivalent magnetic circuit.}
\label{Culasse}
\end{figure}

Ferromagnetic structures \cite{old electromagnet},
which have been develo\-ped \cite{patent} in our group, make possible high gradient fields 
with reasonable power consumption and large condensates. 
Ferromagnetism \cite{Ashcroft} consists in exciting a magnetic material with coils. 
This excitation orientates the microscopic magnetic moments, so that field in the material is enhanced. 
Due to the high permittivity of ferromagnetic materials, the ferromagnetic core guides the magnetic flux.
To understand this effect, let us 
consider the magnetic circuit represented in figure \ref{Culasse}. 
The 
two tips are separated by a gap $e$ of a few centimeters. The 
ferromagnetic structure has a total length $l$ and a section $S\propto r^2$. The 
whole structure is excited by a coil of $N$ loops driven by a 
current $I$, leading to an excitation $NI$. From Ampere's theorem \cite{Jackson}, we can introduce the reluctance $R_{\rm iron}$ :
\begin{equation}
    R_{\rm iron}=\oint_{\Gamma_{\rm i}} \frac{dl}{\mu_{r}\mu_{0}}\simeq\frac{l}{\mu_{r}\mu_{0}}
    \label{eq:1}
\end{equation}
inside the iron core and $R_{\rm gap}$ :
\begin{equation}
    R_{\rm gap}=\oint_{\Gamma_{\rm g}} \frac{dl}{\mu_{0}}\simeq\frac{e}{\mu_{0}}
    \label{eq:1bis}
\end{equation}
in the gap between the tips.
A simple relation between the excitation $NI$ and 
the magnetic flux $BS$ can be written :
\begin{equation}
    BS=\frac{NI}{R_{\rm iron}+R_{\rm gap}}
    \label{eq:2}
\end{equation}
 Since $\mu_{r}$ is very large
($\mu_{r}> 10^{4}$) for ferromagnetic materials, only the gap 
contribution is important ($\mu_{r}\gg l/e$) \cite{foot1}. A more 
complete calculation \cite{DesThes} shows that 
the field created in the gap is similar to that created with two 
coils of excitation $NI/2$ placed close to the tips as represented in figure 
\ref{Culasse}. Thus, guiding of the magnetic field created by 
arbitrary 
large coils far away from the rather small trapping volume is 
achieved. This is only true if a yoke links a north pole to a 
south pole. If not, no guiding occurs and the field in the gap is 
significantly reduced.

Ferromagnetic materials are also subjected to hysteresis cycles generating remnant fields  which must be compensated 
to offer an appropriate magnetic environment. This is somewhat difficult to accomplish as long as some coupling exists 
between the dipole and the quadrupole causing the remnant field to have a complex geometry. In this case, several additional coils are required to provide the right compensating field as previously described in \cite{old electromagnet}. To avoid these couplings, we use a partially-ferromagnetic electromagnet where 
iron cores are only used where strong fields (gradients) are needed (i.e. the quadrupole part of the Ioffe-Pritchard configuration).
With this configuration, we can independantly act on the bias and 
the gradients, and the remnant fields retain a simple geometry.

\section{2. Description and characterization of the electromagnet}

\begin{figure}[!h]
  \centering
  \resizebox{0.9\columnwidth}{!}{\includegraphics{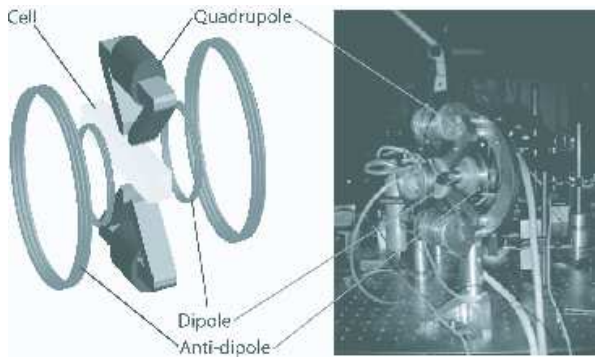}}
  \caption{Scheme and photo of the electromagnet.}\label{FigElectromagnet}
\end{figure}

\begin{figure}[!h]
  \centering
  \resizebox{0.9\columnwidth}{!}{\includegraphics{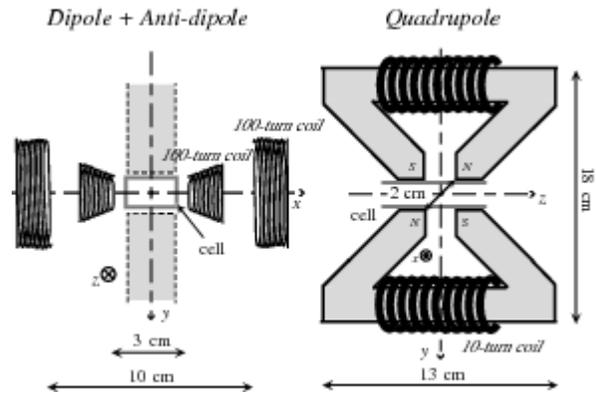}}
  \caption{(a) Configuration of the dipole and anti-dipole coils. (b) Configuration of the quadrupole poles.}
\label{FigPolesAndFields}
\end{figure}

The configuration of the electromagnet is shown in Fig.~\ref{FigElectromagnet} and \ref{FigPolesAndFields}. 
The quadrupole field is produced by two ferromagnetic pieces, each one composing two poles in a way very similar to 
figure \ref{Culasse}. 
Each ferromagnetic piece is excited by a coil of 5~cm diameter made with hollow copper wire (5~mm diameter) 
with 10 windings. The hollow wire allows the exciting coils to be water cooled. Poles are 2~cm wide 
and are bevelled mainly in order to better guide the magnetic field but also to leave some space for the vertical 
magneto-optical trap beams.
The dipolar coils are 100 turn coils of copper wire excited by equal current and separated each 
other by 3~cm. They are conical to facilitate the path of the magneto-optical trap beams and 
positioned on a macor and brass support, which is also water cooled.
Another pair of electromagnetic coils called ``anti-dipole coils" is used to make the bias adjustable. These coils 
are larger than the dipole coils, separated from each other by 10~cm and placed at the support termination. 
The dipole and antidipole coils are mounted in series, but with the currents flowing in opposite directions, so as to cancel out noise in the magnetic field due to current fluctuations.
 We can set the value for the bias field by controling the number of windings of the anti-dipole coils \cite{foot2}. 

\vspace{.5 cm}
\textit{Static performances of the ferromagnetic structure}
\vspace{.5 cm}

\begin{figure}[!h]
  \centering
  \resizebox{0.9\columnwidth}{!}{\includegraphics{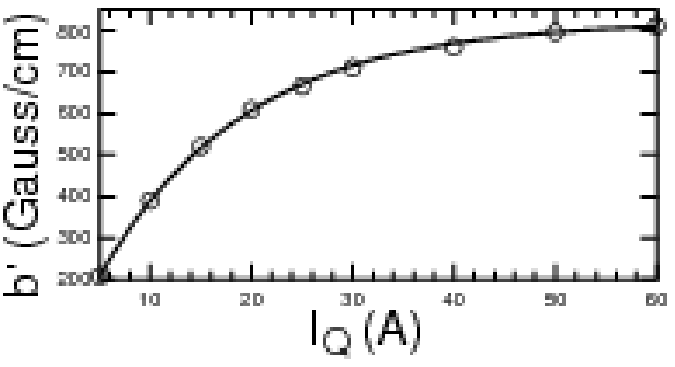}}
  \caption{Magnetic gradient of the quadrupolar field versus intensity $I_Q$.}\label{FigQuadGrad}
  \resizebox{0.9\columnwidth}{!}{\includegraphics{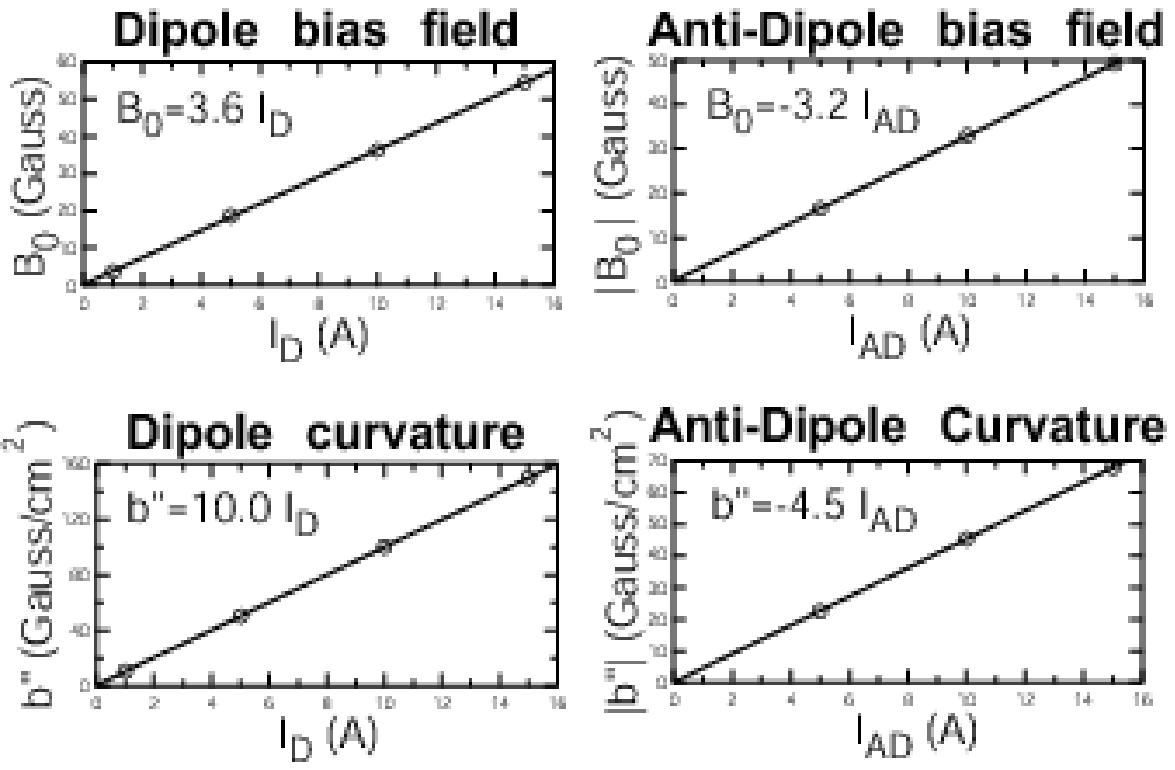}}
  \caption{Bias and curvatures of the fields versus currents $I_D$ and $I_{AD}$ flowing respectively in the dipole and in the anti-dipole.}\label{FigDipPlusAntiDip}
\end{figure}

The maximum achievable quadrupole gradient $b'$ is 830~G/cm, obtained for a saturating current of 60~A (see Fig.~\ref{FigQuadGrad}). 

We limit the current in the dipole coils to 15~A, beyond which heating becomes too important.  With this value, a bias $B_0$ and a curvature $b''$ of respectively 54~G and 150~G/cm$^2$ are obtained for the dipole and 49~G and 68~G/cm$^2$ for the anti-dipole (see Fig.~\ref{FigDipPlusAntiDip}). The total bias can thus be varied from 54 to 5~Gauss.

The modulus of the magnetic field created by the electromagnet is then given by:

\begin{equation}
B=\sqrt{\left(B_0+\frac{b''x^2}{2}\right)^2+\left(b'^2-\frac{b''B_0}{2}\right)\left(y^2+z^2\right)}
\label{chpmag}
\end{equation}

\vspace{.5 cm}

\textit{Dynamical properties}
\vspace{.5 cm}

The dynamical characteristics of the electromagnet are also of a great interest since it is necessary to be able to switch on and off the magnetic field in times shorter than the oscillation period of the  trap. There are two important reasons for this condition. The first one is that we wish to transfer the atoms from the magneto-optical trap to the magnetic trap without losing in density. The second reason is due to the way we image the atomic cloud: it is released from the  trap and expands ballistically before being illuminated  by a near-resonant probe beam (absorption imaging \cite{WKetterle}). From the images, we can deduce temperature, density and the number of trapped atoms.

\begin{figure}[h]
  \centering
  \resizebox{0.9\columnwidth}{!}{\includegraphics{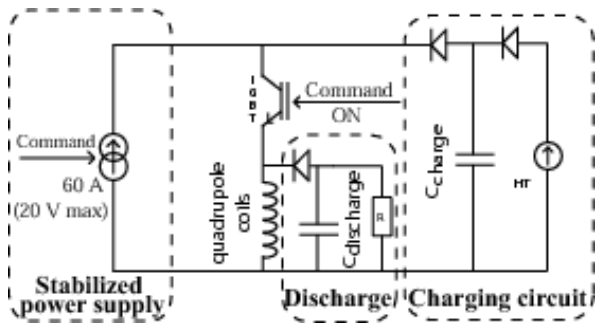}}
  \caption{Electronic scheme of the control of the  current in the quadrupole coils ($L={\rm 1.2 \, mH}$). Switching times are assured by the two capacities with $C_{\rm charge}={\rm 8.8 \, \mu F}$ and $C_{\rm discharge}={\rm 3.14 \, \mu F}$. They are coupled with diodes to limit the oscillation to a quarter of period.} 
\label{FigCircQuad}
\end{figure}

To avoid eddy currents in the ferromagnetic structure, which would limit  switching times, we use a laminated material. It is worth noting that the shape of the present quadrupole is simple enough to be easily produced by assembling inexpensive laser-cut sheets of ferromagnetic material. It consists of 100~$\rm{\mu m}$ thick iron-silicium plates, each isolated from the other.

\begin{figure}[h]
  \centering
  \resizebox{0.9\columnwidth}{!}{\includegraphics{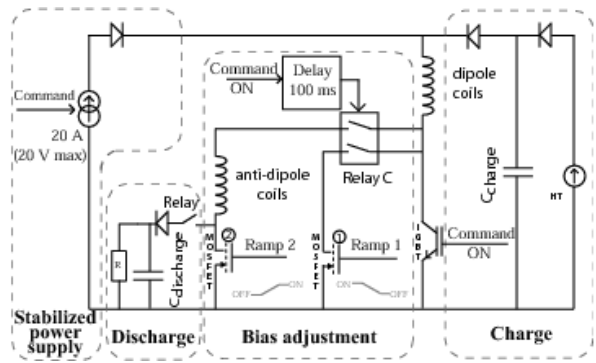}}
  \caption{Electronic scheme of the control of the  current in the dipole ($L={\rm 0.22 \, mH}$) and anti-dipole coils ($L={\rm 0.14 \, mH}$). Switching times are assured by the two capacitors with $C_{\rm charge}=C_{\rm discharge}={\rm 1 \, \mu F}$. They are coupled with diodes to limit the oscillation to a quarter of period. The two MOSFET allow for a control of the current in the anti-dipole coils.}\label{FigCircDip}
\end{figure}

 To control the switching times, we use a capacitor $C$ in series with the coils of inductance $L$. To turn on the magnetic field, the capacitors ($C_{\rm charge}$) are first loaded with a high voltage supply $U$. Then the energy $1/2\,C_{\rm charge}U^2$ accumulated in the capacitors is transferred to the coils in a time equal to $2\pi\sqrt{LC_{\rm charge}}/4$, i.e. a quarter of a period of the oscillating circuit. Then, the power supply takes over. The same idea applies to the procedure to switch off the magnetic field, but now, the switch off current will charge a capacitor called $C_{\rm discharge}$ which limits overvoltage in the circuit at switch off. For our setup, the switching time is  $150~\rm{\mu s}$. The electronic circuit used for the quadrupole is presented in Fig.~\ref{FigCircQuad}. For the dipolar coils (Fig.~\ref{FigCircDip}), the switching procedure is the same, but now we also ramp the current in ``anti-dipole" coils by using  MOSFET so as to precisely adjust the bias at a low value $B_0$ without decreasing the curvature $b''$.  

\vspace{.5 cm}

\textit{Cancelling of the remnant magnetic field }
\vspace{.5 cm}

To generate the compensating field in order to cancel remnant fields, two thin coils (1 mm diameter wire) have been rolled on the two quadrupole exciting coils. They have thus the same configuration and the compensation field automatically possesses the right geometry to cancel the remnant field. The procedure to find the compensation current is quite simple. We adjust the current after having cooled atoms into a magneto-optical trap \cite{Raab} and then cut the MOT magnetic field. If the compensating current is too high or too low, the remnant field provokes a rotation of the atomic cloud, a situation which can be compared to the mechanical Hanle effect \cite{Hanle}. At the right value, there is no rotation anymore, which can be seen by observing fluorescence along the x axis.  

We should stress the point that the remnant fields are so easy to compensate only because we limited the ferromagnetic structure to the quadrupolar plane so as to suppress coupling between quadrupole and dipole. 

\section{3. Atoms trapped in the electromagnet}

\begin{figure}[!h]
  \centering
  \resizebox{0.9\columnwidth}{!}{\includegraphics{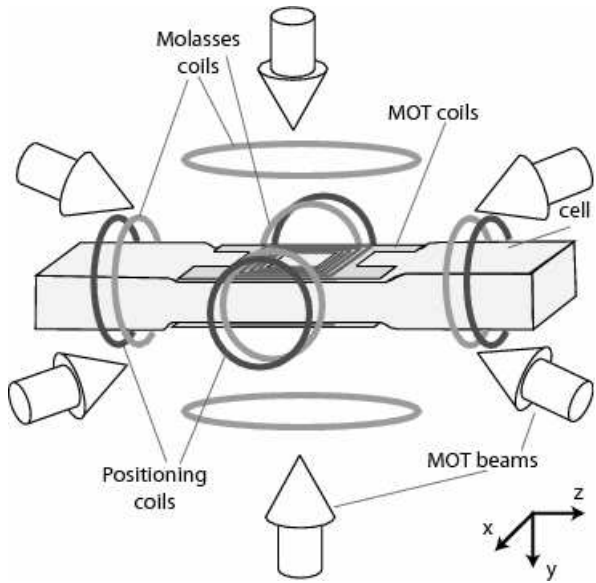}}
  \caption{Configuration of the coils used for the compensation of the external fields (molasses coils in light gray) and the coils used for positioning the MOT (positioning coils in dark grey). Positioning the MOT in the vertical direction is insured by an inbalance in the MOT coils currents.}\label{FigBobines}
\end{figure}

In order to trap the atoms, the electromagnet is placed around a 1 cm square section parallelepipedic glass vacuum cell (see Fig.~\ref{FigBobines}). The atoms are first loaded into a magneto-optical trap whose magnetic field is produced by  two thin printed circuits (400 $\mu$m thick) positioned along the $y$ direction. On each one are imprinted 6.5 loops which are 1 mm wide and 35 micrometers thick, such that the full resistance is 1 $\Omega$.
A current of 2~A in those circuits generates a spherical quadrupole field with an axial gradient of 15~Gauss/cm. A slave laser, injected by a grating stabilized diode laser, produces the six independent circularly polarized trapping beams with a diameter of one centimeter. The MOT sequence ends with an optical molasses \cite{DalCoh} phase during which the  spherical quadrupole field is switched off and the laser frequency is detuned. It is important that there is no residual magnetic field -bias or gradient- during the molasses phase so as not to disturb the sub-doppler cooling mechanism. To ensure this requirement, three pairs of coils (``Molasses coils" in Fig.~\ref{FigBobines}) are placed along the three axes to compensate any residual fields (terrestrial field, ionic pump field...) additional to remnants fields. We obtain $6\times 10^8$ atoms at $50~\mu$K after the molasses.

\begin{figure}[!h]
  \centering
  \resizebox{0.9\columnwidth}{!}{\includegraphics{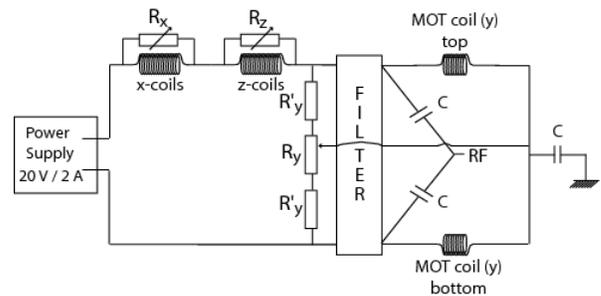}}
  \caption{Electronic scheme for rf circuit and positioning circuit of the MOT relatively to the  magnetic trap.$R_x,R_y{\rm ~and~}R_z$ are variable resistances. The filter cuts frequencies ranging  from 1~MHz to 1~GHz and ensures the proper circulation of the rf signal in the MOT coils. Likewise, capacitors force the current to flow in the opposite direction in the coils.}
\label{FigPositioning}
\end{figure}

After the molasses phase, the atoms are optically transferred into the $F=1$ hyperfine state, 
in which only the $m_F=-1$ sublevel experiences an attractive potential which can be harmonically approximated around the minimum of the magnetic field with the following trapping frequencies:
\begin{eqnarray}
\omega_{\rm ax} & = & \sqrt{\frac{g_F\,m_F\,\mu_B}{m} b''}\\
\omega_{\rm rad} & = & \sqrt{\frac{g_F\,m_F\,\mu_B}{m}\left(\frac{b'^2}{B_0}-\frac{b''}{2}\right)}
\end{eqnarray}
where $g_F=-1/2$ is the Land\'{e} factor and $\mu_B$ the Bohr magneton. 
The values of the trap parameters are presented in table~\ref{TableCaract}.

\begin{table}[!h]
\begin{tabular*}{8.5cm}{l@{\extracolsep{1mm}} *{5}{c @{\extracolsep{2.5mm}}}}
Trap & $B_0{\rm(G)}$ & $b''{\rm(G/cm^2)}$ & $b'{(\rm G/cm)}$ & $\omega_{\rm rad}{\rm (Hz)}$ & $\omega_{\rm ax}{\rm (Hz)}$ \\
\hline
\hline
NCIPT  & 54 & 150 & 200 & 2 $\pi \times$  15 & 2 $\pi \times$ 11 \\
\hline
CIPT  & 5 & 82 & 830 &  2 $\pi \times$ 330 & 2 $\pi \times$ 8  
\end{tabular*}
\caption{Characteristics of the non-compressed trap (NCIPT) and the compressed trap (CIPT). $\omega_{\rm rad}$ and $\omega_{\rm ax}$ are repectively the radial and axial trapping frequencies of the magnetic trap.}
\label{TableCaract}
\end{table}

The magnetic Ioffe-Pritchard trap (IPT) is switched on by applying a current of 15~A in the dipole coils and 5~A in the quadrupole exciting coils. These values are selected so that the size of the two traps -MOT and IPT- are well matched. Moreover, the centers of the two traps have to overlap, otherwise the atoms will acquire a larger energy during the transfer. To adjust the relative position of the trap centers, we have added two pairs of coils in a Helmholtz configuration in series with the MOT's coil. They act on the field in the $x$ and $z$ directions. Variable resistances are added in parallel to the three pairs of coils (including the MOT coils) as shown in Fig.~\ref{FigPositioning}, allowing for tuning of the current in the positioning coils.
The MOT, which will be centered around the zero of the magnetic field, can now be placed appropriately using the positioning coils system described above.

It is to be noted that the right values for each resistance must be found following an iterative method. By moving the center of the MOT, as explained above, the magnetic environment is slightly modified and the molasses coils need to be fine tuned. This in turn displaces the MOT and we have to center the cloud again.

After having transfered the atoms into the IPT, current in the quadrupole is increased to 60~A and the current in the anti-dipole coils is slowly ramped up. At this point, the quadrupole field gradient reaches 830~G/cm and the bias is reduced to 5~Gauss. This provides a tight trap for which the collision rate is around 60~s$^{-1}$, more than one order of magnitude larger than before compression.

\begin{figure}[!h]
  \centering
 \resizebox{0.9\columnwidth}{!}{\includegraphics{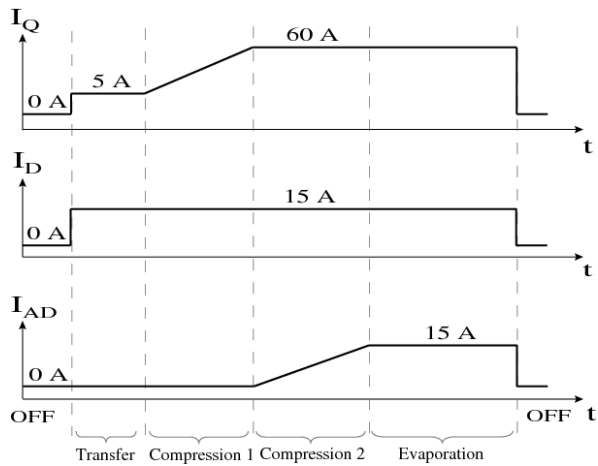}}
  \caption{Temporal sequence for magnetic cycle. $I_Q$, $I_D$ and $I_{AD}$ are currents respectively in Quadrupole, Dipole and Anti-Dipole coils.}\label{FigCycleMag}
\end{figure}

The compression of the trap is adiabatically done in one second in order to minimize the loss of phase space density. Compression time is thus chosen larger than the oscillation time and the rethermalization time. Current values at each stage are summarized in Fig.~\ref{FigCycleMag}. At this stage, we obtain typically $2\times 10^8$ atoms at $250~\mu$K i.e. a phase space density of $\mathrm{5 \times  10^{-7}}$.

\begin{figure}[!h]
  \centering
  \resizebox{0.9\columnwidth}{!}{\includegraphics{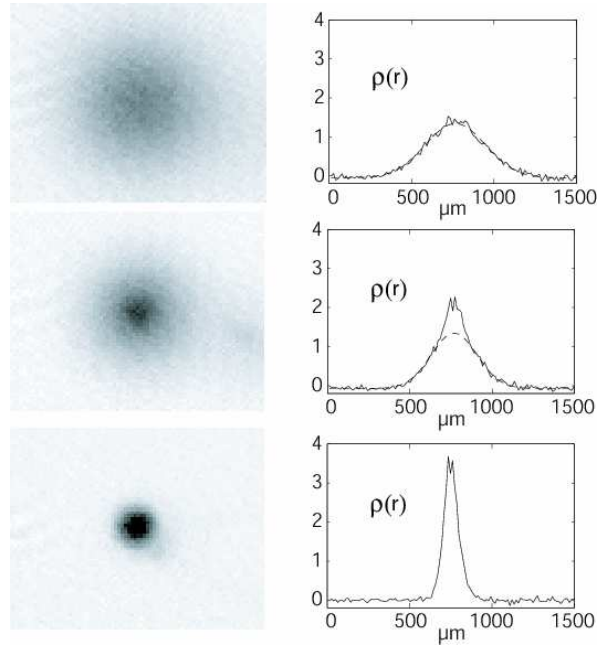}}
  \caption{Atomic cloud, 25 ms after release of the trap, for different final values of the radio-frequency ramp : (a) 4.28 MHz (b) 4.24 MHz (c) 4.20 MHz. $\rho(r)$ is the position dependant atomic density (arbitrary units).}\label{FigTransitionImg}
\end{figure}

Finally, evaporative cooling of the atoms is performed by radio-frequency (rf) induced spin flips. The rf magnetic field is produced by the MOT coils (Fig.~\ref{FigPositioning}), in which the rf current now flows in the same direction for both coils such that the field is nearly homogeneous on the atoms. It is applied perpendicular to the IPT magnetic field at the center of the trap. Then, the radiofrequency is swept from 70~MHz to a final value of around 4.2~MHz. Fig.~\ref{FigTransitionImg} shows images of cooled atomic clouds after 25~ms of ballistic expansion for different final values of the ramp. The sudden appearence of the BEC can be seen in the density profile measured by absorption imaging along the x axis, as the distribution becomes bimodal. 
The distribution of thermal atoms is Gaussian and is represented by a dashed line in the profiles, whereas the distribution of condensed atoms is an inverted parabola (in the Thomas-Fermi approximation)\cite{Stringari}. For decreasing temperatures (lower final rf frequency), the fraction of thermal atoms decreases and at sufficiently low temperatures, we obtain a pure condensate containing $10^6$ atoms.

\section{4. Stability}

\begin{figure}[!h]
  \centering
  \resizebox{0.9\columnwidth}{!}{\includegraphics{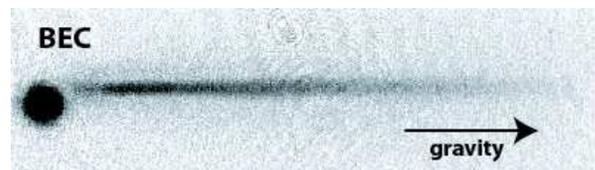}}
  \caption{10 ms of atom laser. The BEC is displaced from the begining of the beam due to a slight Stern-Gerlach effect at the switch off of the trap. The continuity of the laser beam shows qualitatively that the magnetic field is stable}
\label{atomlas}
\end{figure}

\begin{figure}[!h]
  \centering
  \resizebox{0.9\columnwidth}{!}{\includegraphics{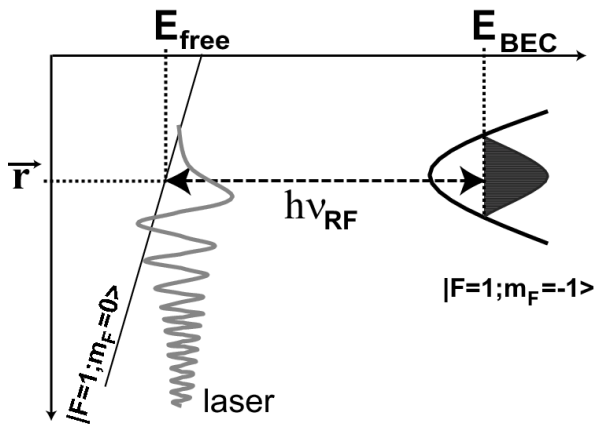}}
  \caption{Principle of radio-frequency coupling of an atom laser}
\label{atomlasprinciple}
\end{figure}

One efficient way to check the stability of the bias created by the electromagnet is to produce atom lasers by radio-frequency outcoupling \cite{Bloch99}(Fig. \ref{atomlas}) and mesure the spectral width of the condensate.
The principle of weak radio-frequency outcoupled atom lasers has been developped in \cite{paplaser}.
Main features are summarized in figure \ref{atomlasprinciple}.
The condensate is displaced from the center of the magnetic field due to gravity. Thus, it crosses magnetic equipotentials in the vertical direction only.

By applying radio-frequency one flips the spin of atoms from the trapped state $|F=1,m_F=-1>$ to the untrapped one $|F=1,m_F=0>$. The wave falls under the effect of gravity into the continuum of Airy functions \cite{Landau}. This gives rise to an atom laser since the source (the condensate) is fully coherent \cite{cohbec}. The radio-frequency  value $\nu_{RF}$  is directly related to the difference in energy between the two involved states:
$$
h\nu_{RF}=E_{\rm BEC}-E_{\rm free}(\vec{r})=E_{\rm bias}+\mu-E_{\rm free}(\vec{r})
$$ 
where $E_{ \rm free}(\vec{r})$ stands for the potential energy of outcoupled atoms and depends on the extrating point $\vec{r}$, $\mu$ and  $E_{\rm bias}$, respectively the chemical potential and  the magnetic potential energy of the BEC. For a given radiofrequency, fluctuations of the coupling zone are only related to bias fluctuations through $E_{\rm bias}$.

\begin{figure}[!h]
  \centering
  \resizebox{0.9\columnwidth}{!}{\includegraphics{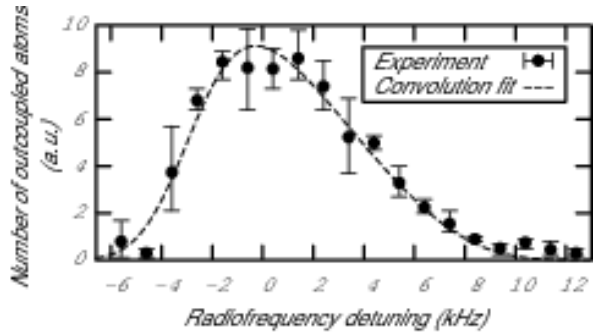}}
  \caption{Comparison between the experimental width and a Thomas-Fermi width broadened by a bias instability of 2~mG. The curve is clearly assymetric due to the fact that the radiofrequency knife has a strong curvature since the gravitational displacement $g/\omega_{\rm rad}^2=2.3 ~\mu m$ is shorter than the Thomas-Fermi radius $R_y=3 ~\mu m$. }
\label{width}
\end{figure}

To quantitavely give an estimation on the stability of the experiment, we have to determine the spectral width of the condensate. For a given weak radio-frequency power, we have measured the number of extracted atoms as a function of the value of the rf knife (experimental points in figure \ref{width}). If we assume that the number $N_e$ of extracted atoms is proportional to the number of condensed atoms on the surface $\Gamma(\delta\nu)$ defined by the rf value $\nu_{RF}$ (i.e. one neglects the effect of interactions on the coupling rate \cite{paplaser}). Taking a Thomas-Fermi profile for the condensate leads to:

$$
N_e(\delta\nu)\propto \int_{\Gamma(\delta\nu) }  \,d \vec{x} \,{\rm max}\left(1-\sum_{i=x,y,z}\left(\frac{x_i}{R_i}\right)^2;0\right)
$$
where $\delta\nu=\nu_{RF}-\nu_0$ is the difference between outcoupling radio-frequency and its value in the center of the BEC, $x_i$ and $R_i$ being respectively the coordinate and the Thomas-Fermi radius of the dimension $i$. The surface $\Gamma(\delta\nu)$ corresponding to the radiofrequency knife resonance condition is defined by:
$$
\omega_{\rm ax}^2 x^2 +\omega_{\rm rad}^2\left(y^2+z^2\right) + 2gy=\frac{2h}{m}\delta\nu
$$
 
If we assume that fluctuations are gaussian with r.m.s value $\sigma$, the final output flux is the convolution between this noise and $N_e$:

$$
F(\delta \nu)=\int du \,N_e(u) \,e^{-(u-\delta\nu)^2/2\sigma^2}
$$

We have fitted the experimental data with $F$ as shown in figure \ref{width}, and we obtain $\sigma=$1.5~kHz. This broadening could come from bias variations during one scan.

In order to look more precisely for the different timescales involved in these fluctuations, we have fixed the rf outcoupler at a detuning of $\mathrm{+5\, kHz}$ and we have looked at the long term drift and fluctuations of the number of outcoupled atoms. Once we have deduced the effects of atom number fluctuations in the original BEC ($\mathrm{8\,\%}$), we have measured a drift of $\mathrm{500 \,Hz/hour} \,(\mathrm{0.7 \,mG/hour})$ and r.m.s. fluctuations of $\mathrm{700\, Hz}$ which correspond to shot to shot bias fluctuations of $\mathrm{1\, mG}$.

Nevertheless, it is likely that the rapid fluctuations are also of the order of $\mathrm{1\,mG}$, which is sufficient to produce stable quasi-continuous atom lasers as one can see on figure \ref{atomlas}.

\section{Conclusion}

In this paper, we have presented the realization of a new generation hybrid electromagnet designed to produce large Bose-Einstein condensates containing one million Rubidium atoms. Due to the compact ferromagnetic design (20~cm diameter) one achieves gradients as large as 830~G/cm which are sufficient to compress the atomic cloud for an efficient runaway evaporative cooling. The total power consumption is less than 800~W which is easy to dissipate. Due to the lamellar structure of the ferromagnetic core and the driving electronic circuit, eddy currents are suppressed and the switching time is reduced to 150~$\mu$s. The high stability of the bias (fluctuations measured to be of the order of $\mathrm{1 \, mG}$) enables us to produce stable radio-frequency outcoupled quasi-continuous atom lasers.

\begin{acknowledgments}
 Y. Le Coq acknowledges support from the CNES post-doctoral fellowship program. This work is part of the CNES supported ICE project (DA No. 10030054) with initial support from Laboratoire National d'Essai, D\'{e}l\'{e}gation G\'{e}n\'{e}rale de l'Armement (Contract No. 99-34-050), the European Union (Cold Quantum Gas network) and INTAS (Contract No. 211-855). We would like to thank John Gaebler for his careful reading of this manuscript. 
\end{acknowledgments}


\begin{thebibliography}{llml}
\bibitem{foot1} The case of very small gap where 
$R_{iron}\geq R_{gap}$ was studied in \cite{Vuletic96}. In this case, the 
ferromagnetic materials amplify the magnetic field in the gap.
\bibitem{foot2} It is worth noting that having an adjustable bias is not simple in the 
 case of an entire ferromagnetic structure : the anti-dipole field would couple into the ferromagnetic structure, 
 which would lead to a desexcitation of the material and would reduce the longitudinal curvature too much.
\end{thebibliography}
\end{document}